\documentclass[journal]{vgtc}                     

\onlineid{0}

\vgtccategory{Research}

\vgtcpapertype{theory/model}

\title{The Language of Infographics: Toward Understanding\\ Conceptual Metaphor Use in Scientific Storytelling}

\author{%
  \authororcid{Hana Pokojná}{0000-0001-7854-2596}, 
  \authororcid{Tobias Isenberg}{0000-0001-7953-8644},
  \authororcid{Stefan Bruckner}{0000-0002-0885-8402},
  \authororcid{Barbora Kozlíková}{0000-0003-0045-0872},
  \authororcid{Laura Garrison}{0000-0001-7134-2006}%
  }

\authorfooter{
  \item
  	Hana Pokojná is with Visitlab, Faculty of Informatics, Masaryk University, Czech Republic.
  	E-mail: hpokojna@mail.muni.cz.
  \item
  	Tobias Isenberg is with Inria, Université Paris-Saclay, CNRS, Inria, LISN, France.
  	E-mail: given\_name.family\_name@inria.fr.

  \item Stefan Bruckner is with the Institute for Visual and Analytic Computing, University of Rostock, Germany.
  	E-mail: stefan.bruckner@uni-rostock.de.
   
   \item Barbora Kozlíková is with Visitlab, Faculty of Informatics, Masaryk University, Czech Republic.
  	E-mail: kozlikova@mail.muni.cz.
   
   \item Laura Garrison is with the Department of Informatics \& Mohn Medical Imaging and Visualization Centre, Department of Radiology, Haukeland University Hospital, University of Bergen.
  	E-mail: laura.garrison@uib.no.
}

\abstract{%
We apply an approach from cognitive linguistics by mapping Conceptual Metaphor Theory (CMT) to the visualization domain to address patterns of visual conceptual metaphors that are often used in science infographics. 
Metaphors play an essential part in visual communication and are frequently employed to explain complex concepts. However, their use is often based on intuition, rather than following a formal process.   
At present, we lack tools and language for understanding and describing metaphor use in visualization to the extent where taxonomy and grammar could guide the creation of visual components, e.g., infographics.
Our classification of the visual conceptual mappings within scientific representations is based on the breakdown of visual components in existing scientific infographics. 
We demonstrate the development of this mapping through a detailed analysis of data collected from four domains (biomedicine, climate, space, and anthropology) that represent a diverse range of visual conceptual metaphors used in the visual communication of science.
This work allows us to identify patterns of visual conceptual metaphor use within the domains, resolve ambiguities about why specific conceptual metaphors are used, and develop a better overall understanding of visual metaphor use in scientific infographics. 
Our analysis shows that ontological and orientational conceptual metaphors are the most widely applied to translate complex scientific concepts. To support our findings we developed a visual exploratory tool based on the collected database that places the individual infographics on a spatio-temporal scale and illustrates the breakdown of visual conceptual metaphors. 
}


\teaser{
  \centering
  \includegraphics[width=\linewidth]{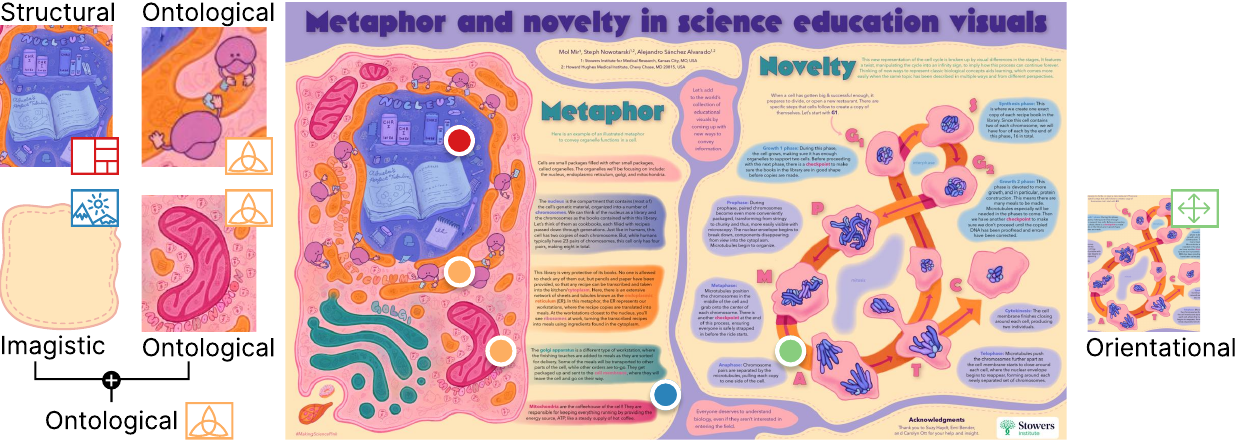}
  \caption{Example of a deconstructed science infographic from the biomedical domain \cite{VIZBI2023}, with each conceptual metaphorical element classified as a structural, ontological, orientational, or imagistic metaphor. 
  An \emph{imagistic} metaphor correlates one concept with another employing graphical similarity. 
  Here, the wobbly structure of a cell membrane is graphically described through an irregular patch shape inscribed with stitch marks. 
  An \emph{orientational} metaphor describes a concept (or system of concepts) based on spatial orientation and composition, e.g., the cell cycle where its stages are arranged in an infinity symbol. 
  An \emph{ontological} metaphor projects concrete entities onto abstract concepts (e.g., events, actions, states, activities), e.g., presenting ribosomes as cafe workers. 
  A more complex \emph{structural} metaphor ties one unfamiliar conceptual network or set of processes to another more familiar conceptual network through an entailment relationship, e.g., the properties and functions of the nucleus are represented as a library. 
  Conceptual metaphors may be combined to leverage greater meaning, such as combining the wobbly patch shape with the mitochondrial coffeehouse nested therein to an ontological metaphor for a \textit{cellular} energy source.
  }
  \label{fig:teaser}
}

\graphicspath{{figs/}{figures/}{pictures/}{images/}{./}} 

\usepackage{tabu}                      
\usepackage{booktabs}                  
\usepackage{lipsum}                    
\usepackage{mwe}                       

\usepackage{mathptmx}                  

\usepackage{graphicx}

\usepackage{color, colortbl}

\usepackage[bookmarks]{hyperref}

\usepackage{pifont}

\usepackage{amssymb}

\usepackage{amsmath}
\usepackage{wasysym}

\usepackage[normalem]{ulem}

\usepackage[dvipsnames,svgnames]{xcolor}
\usepackage{amssymb}

\usepackage{graphicx,calc}
\newlength\myheight
\newlength\mydepth
\settototalheight\myheight{Xygp}
\newcommand*\inlinegraphics[1]{%
  \settototalheight\myheight{Xygp}%
  \settodepth\mydepth{Xygp}%
  \raisebox{-\mydepth}{\includegraphics[height=\myheight]{#1}}%
}

\usepackage{ccicons}
\usepackage{xspace}

\newcommand{\osflink}[1]{\href{https://osf.io/8xrjm/}{#1}}
\newcommand{\osflinkrepo}{\osflink{\texttt{osf\discretionary{}{.}{.}io\discretionary{/}{}{/}8xrjm}}\xspace}

\newcommand{\toollink}[1]{\href{https://osf.io/hvfp3}{#1}}

\newcommand{\databaselink}[1]{\href{https://osf.io/pbyhs}{#1}}

\newcommand{\figureslink}[1]{\href{https://osf.io/r7x6e}{#1}}

\newcommand{\keywordslink}[1]{\href{https://osf.io/3gpqw}{#1}}

\newcommand{\livetoollink}[1]{\href{https://lauragarrison87.github.io/metaphorTool/exploratoryTool/}{#1}}

\newcommand{\githubrepolink}[1]{\href{https://github.com/lauragarrison87/metaphorTool}{#1}}

\newcommand{\humanmigrationmaplink}[1]{\href{https://www.migrationheritage.nsw.gov.au/objects-through-time/essays/50000-years-before-present/attachment/human-migration-map/index.html}{#1}}

\newcommand{\bigbanglink}[1]{\href{https://en.wikipedia.org/wiki/Big_Bang}{#1}}

\newcommand{\humanevolutionlink}[1]{\href{https://www.visualcapitalist.com/path-of-human-evolution/}{#1}}

\newcommand{\stateoftheuklink}[1]{\href{https://www.rmets.org/news/climate-change-continues-be-evident-across-uk}{#1}}

\newcommand{\wildfireslink}[1]{\href{https://www.ucsusa.org/resources/western-wildfires-and-climate-change}{#1}}

\newcommand{\worldhistoryencyclopedialink}[1]{\href{https://www.worldhistory.org/}{#1}}

\newcommand{\migrationperiodlink}[1]{\href{https://www.worldhistory.org/image/14250/migration-period-in-europe-during-the-4th--5th-cen/}{#1}}

\begin{document}


\firstsection{Introduction}

\maketitle


The philosopher Ludwig  Wittgenstein once declared that ``the limits of my language mean the limits of my world,'' \cite{wittgenstein2023tractatus} referring to the constraints that spoken words and language impose on human capacity to understand and engage with the world. Put another way, if a feeling or similarly abstract experience cannot be described, then this feeling or experience \textit{does not exist}. To convey abstract realities and expand our knowledge of reality, we often use metaphors---familiar and conceptually similar or comparable descriptors for unfamiliar, possibly intangible, concepts or events. 

Metaphors play an important role in information transfer, e.g., in education \cite{low2008metaphor} and medical communication \cite{sanchez2018visual, preim2023survey, ferrando2017conceptual}. Their extensive study in the linguistics domain \cite{searle1969speech, lakoff1993contemporary,davidson1979metaphors-alternative} has led to the development of systematic approaches for communication that can enable deeper reasoning about visualizations as communication devices. 
Their ability to clarify unfamiliar concepts bridges disciplines by establishing a shared conceptual foundation, for instance, presenting an unfamiliar concept like the cell's nucleus and its functions as a familiar concept of a library with its related tasks (\cref{fig:teaser}). Effective science communication follows many rules \cite{jensen2006metaphors}, and often relies on metaphors considered essential for understanding scientific findings \cite{brown2003making}. Previous research points out the benefits of visual components in addition to text. For instance, the Multimedia Learning Theory (MLT) \cite{mayer1998cognitive} states that readers understand concepts better from text accompanied by images. This effect was also demonstrated by improved memorization and response rates for study participants provided with visual embellishments \cite{borgo2012empirical} or infographics in an educational setting \cite{parveen2021infographics}.

Our investigation into metaphor (de)construction is inspired in part by Robert Hooke's 1665 publication \emph{Micrographia} \cite{hooke1968micrographia}, a meticulously described collection of microscopic structures, scientific visualizations, and guidelines for using the newly-developed microscope. Through a new visual vocabulary steeped in metaphor, Hooke taught readers \textit{what} to see, as well as \textit{how} to see and appreciate this new microscropic world and its inhabitants while onboarding them to a rigorous scientific process of empirical observation \cite{jack2009pedagogy}.  
%
Part of \emph{Micrographia}`s genius is Hooke`s leveraging of collective known experiences to encode understanding of the unfamiliar in a so-called \textit{image schema}, ``a recurring, dynamic pattern of our perceptual interactions and motor programs that gives coherence and structure to our experience'' \cite{johnson2013body}. In essence, these schemas are conceptual frameworks that facilitate knowledge gain by mapping known familiar experiences to the unknown, thereby lowering cognitive load \cite{SWELLER1988257}. In visualization, we see such schemas applied to information transfer and understanding of data \cite{li2017impact}, e.g., through structuring diagrams \cite{nguyen2014unlocking} or representational icons to facilitate learning \cite{ziemkiewicz2008shaping} and memorization \cite{borgo2012empirical}. Metaphorical visual mappings may furthermore lead to the extension and reanalysis of an existing schema \cite{fauconnier1997mappings}, enabling the designer to leverage a particular visual element to support their rhetorical aim(s) \cite{hullman2011visualization}. These schemas can also be manipulated to tell a powerful story through \emph{affective} visualization that ``relates to, arises from, or results in emotion''\cite{lan2023affective}, by which the visual sheds light on particular attributes that guide the viewers to react in a certain way \cite{xiao2021we} or highlights the importance of their actions and consequences \cite{samsel2021affective}, e.g., through strategic use of color semantics \cite{schloss2023color, lin:2013:selecting}.

 Well-established theories from other domains may help us better understand the full potential and implications of using metaphors in visualization. Conceptual Metaphor Theory (CMT), first applied to the analysis of literary language \cite{lakoff1980metaphors}, coins the term ``conceptual metaphors'' and their use as a matter of thought and reasoning extending beyond linguistics. Conceptual metaphors are widely used to imbue meaning and are often at the center of abstract conceptualization \cite{landau2010metaphor}. 
 CMT describes four major conceptual metaphors with varying conceptual complexity: \emph{imagistic}, \emph{orientational}, \emph{ontological}, and \emph{structural} (\cref{fig:glyphs}), also discussed by Vu \cite{vu2015structural}. While CMT has previously provided the theoretical underpinnings for visual representation choices \cite{barr2005taxonomic,ziemkiewicz2008shaping, preim2023survey, risch2008role, tkachev2022metaphorical}, there is little exploration of the types of metaphors CMT describes. 

Visualization combines human-- and machine--centric approaches inspired by and rooted in other fields such as psychology, mathematics, and neuroscience \cite{chen2017pathways}, and can be further enriched by linguistics as this field informs visual metaphor construction and use. Such work, in turn, can further elucidate other scientific disciplines through appropriate data depiction using suitable visual conceptual metaphors. 
Our work applies CMT to scientific infographics and discusses findings for how different types of conceptual metaphors may help to disambiguate the graphical communication of science. 
Metaphor use has a degree of subjectivity and can vary considerably across different cultures, societies, and even individuals \cite{lakoff1980metaphors, hullman2011visualization, kitao1986study}. We limit our study to a predominantly Western, English-speaking context, where all co-authors are also situated. 
Specifically, we deconstruct and analyze visual conceptual metaphors in scientific infographics based on the established linguistic Method for Identifying Metaphorically Used Words (MIP) \cite{group2007mip} and CMT. We observe trends, impacts, and implications of metaphor-type usage in four scientific domains: biomedicine, climate, space, and anthropology.
In summary, we contribute:
\begin{itemize}[nosep,left=0pt .. \parindent]
\item a discussion of conceptual metaphor types and their classification, grounded in CMT, in visual infographics across four scientific domains,
\item a database of classified conceptual metaphors in scientific infographics to demonstrate how visual conceptual metaphors are used in practice, and
\item a visual tool for the exploration of conceptual metaphor use for each infographic within our collected corpus, organized by scientific domain and the spatio-temporal range of processes depicted with each infographic.
\end{itemize}

\section{Core definitions and concepts}
\label{sec:define}
\begin{figure}[t]
        \centering 
         \includegraphics[width=0.9\columnwidth]{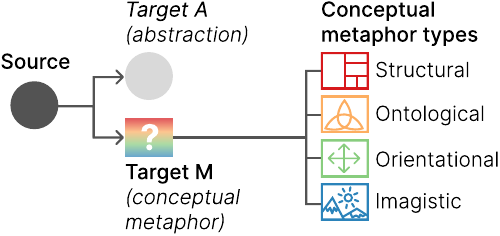}\vspace{-1ex}
         \caption{Semiotic depiction of the target and source domains and the four types of conceptual metaphors used in this work. Mapping from a source to target in the same conceptual domain (Target A) produces an abstraction. Mapping from a source to target in a \textit{different} conceptual domain (Target M) produces an conceptual metaphor that can be further identified as a \emph{imagistic, orientational, ontological,} or \emph{structural} metaphor.%
         }
         \label{fig:glyphs}
    \end{figure}

Our first contribution in this work is a detailed discussion of what does and does \textit{not} qualify as a conceptual metaphor, followed by the four types of conceptual metaphors on which our analysis centers (\cref{fig:glyphs}), as well as other important properties to consider in metaphoric depictions. 

\subsection{Conceptual metaphor identification}
\label{conc_met}

 \begin{figure}[b]
        \centering 
         \includegraphics[width=0.67\columnwidth]{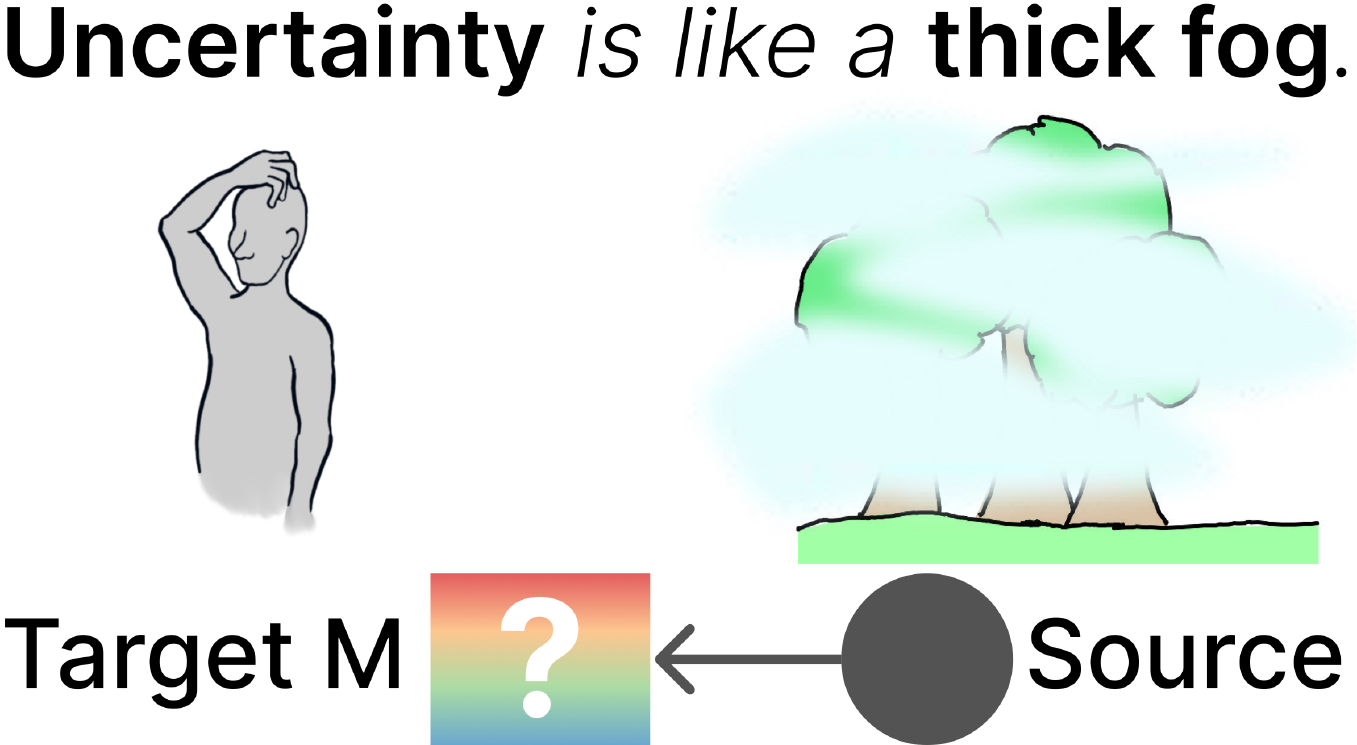}\vspace{-1ex}
         \caption{An example visual conceptual metaphor of \textbf{fog} for the notion of \textbf{uncertainty}. Uncertainty is an abstract, unfamiliar concept (Target M), graphically mapped to the familiar, more concrete concept of fog shrouding a group of trees (Source). 
         }
         \label{fig:mapping}
    \end{figure}
    
Lakoff and Johnson \cite{lakoff1980metaphors} define a \textbf{conceptual metaphor} as a way of describing concepts from an \textit{unfamiliar (target)} domain using the concept(s) from a more \textit{familiar (source)} domain, resulting in a \textbf{cross-do\-main} mapping \cite{lakoff2006conceptual}. Conceptual metaphors imply relationships between the source and target domains, usually by mapping familiar onto unfamiliar ideas to relate shared perceptual experiences and to enable more complex reasoning \cite{pinker2007stuff}. A conceptual metaphor mapping is typically unidirectional \cite{lakoff1980metaphors} and moves from a concrete or physical source to a more abstract target \cite{lakoff1980metaphors,gentner1997alignment,boroditsky2000metaphoric} (\cref{fig:glyphs}), although this mapping may also be bidirectional \cite{huangrepresent, porat2017metaphor,danesi2017bidirectionality,lee2012bidirectionality}. 
%
As previously described, metaphors extend beyond linguistics and textual embellishments \cite{jones1992generating} to shape thought, reasoning, and cognition \cite{lakoff1980metaphors, lakoff1993contemporary}. In the context of visualization, therefore, a graphical representation from a more familiar (source) domain, such as fog, to explain a more abstract (target) domain, such as uncertainty, creates a \textbf{visual conceptual metaphor (\cref{fig:mapping}).} 

\begin{figure}[tb]
        \centering 
         \includegraphics[width=0.9\columnwidth]{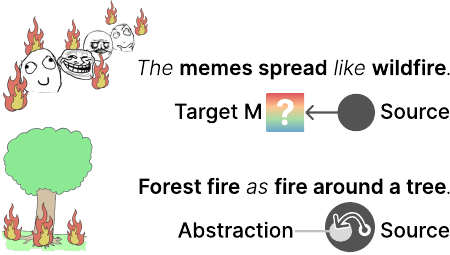}
         \caption{%
         \textit{\textbf{Top:}} a conceptual metaphor, visually depicting the rapid popularity of \textbf{memes spreading} (Target M) as quickly as a \textbf{wildfire} (Source). The concept of physical wildfire and its fast-spreading property describes the concept of rapid popularity increase and usage. 
         \textit{\textbf{Bottom:}} visual depiction of an abstraction, a graphic of \textbf{fire around a tree} (Abstraction) to denote the concept of a \textbf{forest fire} (Source). This does not provide more understanding of the forest fire beyond graphically depicting the event. Glyph explanations are inspired by Szefliński \& Wachowski \cite{szeflinski2019metaphor}.}
         \label{fig:abstractionVsMetaphor}
    \end{figure}

Conceptual metaphorical depiction occurs when the target and source domains \emph{differ}. Based on this definition, a \textbf{visual abstraction} of a concept is not a metaphor, only a simplified depiction of a target mapped \textit{within the same domain} as the source. 
We note that this definition of abstraction differs from the usual notion of abstraction in visualization, whereby a visual is reduced to its essentials with unnecessary details removed \cite{Viola:2018:PCA,Viola:2020:VA,Wu:2023:TFA}.
Instead, a linguistic example, \textbf{metonymy}, is better-aligned our definition of visual abstraction of concepts or ideas. In contrast to a conceptual metaphor, metonymy maps an aspect of the target domain to describe the target in its entirety, e.g., \emph{the buses are on strike.} \cite{lakoff1980metaphors}. Here, \textbf{buses} are an abstraction for the \textbf{drivers of the buses}, who are the entities on strike. Similarly, an image of a fire can be a visual abstraction for the notion of a forest fire (\cref{fig:abstractionVsMetaphor}, bottom). We contrast this with a conceptual metaphor (\cref{fig:abstractionVsMetaphor}, top), which crosses conceptual domains to depict the spreading of memes (target domain) as a wildfire (source domain).

\subsection{Types of conceptual metaphors}
\label{met_types}
Conceptual metaphors can further be classified as \emph{imagistic}, \emph{orientational}, \emph{ontological}, and \emph{structural} \cite{lakoff1980metaphors, macha2016conceptual}. 
Below, we refer to \cref{fig:teaser} as a running example for each type of visual conceptual metaphor. 


\textbf{Imagistic. \inlinegraphics{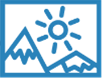}}  
An \emph{imagistic} metaphor correlates one concept with another employing \emph{graphical similarity}; it is the simplest of the four types of metaphors. It creates an image in the reader's head that describes the source concept but does not convey any deeper meaning or functional properties. In the cell infographic in \cref{fig:teaser}, the membrane of a cell (target) is metaphorically visualized by an irregularly-shaped patch inscribed with stitch marks around its interior (source). This image provides a sense of appearance without providing more information about its properties or functions. The cell could alternatively represented by a simple circle or ellipse---but that would be an abstraction. 

\textbf{Orientational. \inlinegraphics{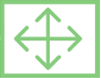}}
An \emph{orientational} metaphor describes a concept or system of concepts based on \emph{spatial orientation and composition}. An example is a continuous repeating life cycle in \cref{fig:teaser}, where the temporal properties---cyclicality and repetitiveness (target)---are described by strategically arranged stages in a pattern of an infinity symbol (source). The spatial arrangement, or positioning, of visual sub-elements map the concept of repetition to the cellular life cycle (target). An orientational metaphor can also denote affective properties, e.g., positive and negative concepts, through positioning of elements on the canvas: in Western, English-speaking societies, \emph{up} typically indicates positive, while \emph{down} often has negative connotations \cite{meier2004sunny}. Moreover, such alignment of visual position with emotion may also improve readability \cite{woodin2021conceptual}.

\textbf{Ontological. \inlinegraphics{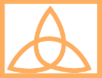}} 
An \emph{ontological} metaphor \emph{projects concrete entities onto abstract concepts} (e.g., events, actions, states, activities). Through this projection, the abstract concepts are given `borders' \cite{lakoff1980metaphors} and thus can then be referred to and quantified as concrete objects. 
For instance, the concept of \textit{the \textbf{mind} as a \textbf{machine}} facilitates the discussion of the abstract notion of the human mind through the more tangible language of the workings of a machine, e.g., \emph{the mind becoming rusty} or \emph{the mind not operating}~\cite{lakoff1980metaphors}. In visualization, color semantics can be leveraged as ontological metaphors, e.g., red to convey situational urgency (\cref{fig:infographics_2}b). Personification is another projection strategy in ontological metaphors, which enables the discussion of abstract concepts or ideas through the most relatable source domain possible---ourselves \cite{kovecses2010metaphor}. For instance, in \cref{fig:teaser} the ribosomes (target) in a cell are referred to as cafe workers (source) due to their function (transcription). This type of metaphor requires the viewer to think more abstractly to associate a physical entity or a substance (source) with its function or semantic meaning (target) through the lens of human activities.

\textbf{Structural. \inlinegraphics{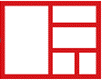}} 
A \emph{structural} metaphor is the most complex of the four metaphors; it maps an entire network of concepts from a source onto the unfamiliar network in the target domain. Lakoff and Johnson \cite{lakoff1980metaphors} provide the literary example \textit{\textbf{argument}} (target) \textit{is} \textit{\textbf{war}} (source). This metaphor uses the highest level of conceptual abstraction to explain a set of concepts (i.e., \textit{argument} involving people, unpleasant feelings, time, strategy, and actions of attacking or defending a point) through a related but conceptually different set of processes that involve negative feelings, effort, time, strategy, and actions of attacking or defending, etc.).
This type of systematic framing of a concept is achieved through entailment linking between processes. In \cref{fig:teaser}, e.g., the cell nucleus (target) being visually represented as a library (source) conveys the idea that this organelle serves as a library and holds books (information) that are organized in a specific way, where each book can be borrowed, read, and copied. The library concept entails that the nucleus has its properties, uses, and functions: holding information that is organized in a specific way and can be taken out, read, and transcribed. This concatenation of familiar processes (library) paints a picture of another set of lesser-known and more conceptually abstract processes. 

Structural metaphors may be, in some cases, difficult to tell apart from \emph{ontological} metaphors, such as when the ontological metaphors are grouped with other conceptual metaphors, e.g., \emph{imagistic}, to create a more conceptually rich \emph{ontological} metaphor. Additionally, per Lakoff and Johnson \cite{lakoff1980metaphors}, structural metaphors are themselves may result from the layering or nesting of two or more conceptual metaphors. Consider the structural metaphor \textit{\textbf{time} is a \textbf{resource}} (\cref{met_types}): \emph{time} is referred to as an object that may be quantified (i.e., is made an ontological metaphor) and its identification as a \emph{resource} is similarly an ontological metaphor for the limits or preciousness of time which entails the notion of time as an object. Taken together, this structural metaphor denotes time as a countable, limited, and precious commodity. In contrast, an \emph{ontological} metaphor on its own differs by limiting representation to a singular concept absent an extended network of meaning. 

\subsection{Additional conceptual metaphor attributes}
In addition to understanding the basic concepts, the following attributes provide more insights into how metaphors can be structured and understood. 

\textbf{Incidence hierarchy.}
Lakoff and Johnson \cite{lakoff1980metaphors} note that some conceptual metaphors are more prominent and recognizable than others, noting that structural metaphors in particular are pervasive in thought, e.g., \textit{\textbf{time} is \textbf{money}} or \textit{\textbf{love} is a \textbf{journey}}. Culture naturally impacts interpretation and use of certain types of metaphors. \emph{Time is money} carries different connotations in monastic subcultures that emphasize a value system of \emph{less is more} \cite{lakoff1980metaphors}. Individual perspectives and cultural trends may furthermore influence the incidence of certain metaphors, e.g., \emph{climbing the corporate ladder} is an orientational metaphor that may be viewed positively (a motivated individual) or negatively (a cutthroat individual) within a larger sociocultural and temporal context. Some metaphors may be so pervasive as to transcend culture: a humanoid robot is widely recognized across contemporary modern societies as a metaphor for artificial intelligence (AI). This metaphor influences reasoning and actions related AI, insinuating human rationality and intelligence where only stochastic processes exist \cite{hermann2023artificial}. 


\textbf{Complexity hierarchy and conceptual nesting.}
While Lakoff and Johnson \cite{lakoff1980metaphors} do not explicitly state that conceptual metaphors are hierarchically organized based on complexity, a study of metaphors in visualization by Ziemkiewicz and Kosara \cite{ziemkiewicz2008shaping} asserts that there are assumptions about how visualizations are structured, as well as how they interrelate and can be broken down into sub-elements. These assumptions mean that visual metaphors contribute to the understanding of the presented information and that organizational changes in these metaphors can change their meaning \cite{ziemkiewicz2008shaping, ziemkiewicz2009preconceptions, ziemkiewicz2010understanding}. Kövescses~\cite{kovecses2010metaphor} suggests a hierarchy in conceptual metaphors based on their level of abstraction and that their combination can create more complex metaphors, thus creating a multilevel schematicity. 
%
As previously discussed in \cref{met_types} through the \emph{time as a resource} example, structural metaphors may often leverage such a combination of conceptual metaphors.
This joining or nesting of individual metaphors into more complex ideas is also evident in some infographics. In \cref{fig:teaser}, the visual conceptual metaphor for a cell membrane (\emph{imagistic}) as a wobbly stitched patch is not conceptually divisible. It can be nested, however, into an \emph{ontological} metaphor where the patch element contains more units with specific functions---like a coffeehouse as a mitochondrion to say that the mitochondrion provides energy for the cell. Using hierarchical or nesting strategies to build more complex metaphors may not always be possible, and would benefit from deeper investigation into the true extent of such strategies that falls beyond the scope of this work. 



\section{Related work}
Our investigation into CMT as a conceivable guiding principle for analyzing science infographics and visualizations is rooted in diverse perspectives in visualization research and practice, considering epistemologies of visualization and metaphors as visual rhetorical objects.

\textbf{Epistemologies of visualization.}
Visualization research and practice draw from and build upon a diversity of disciplines that, in turn, provide different lenses for ways of knowing and reasoning about visualization. Chen et al.~\cite{chen2017pathways} argue for developing a stronger theoretic foundation in visualization, including the advancement of taxonomies and ontologies---our work moves in this direction. Van Wijk~\cite{van2005value} contemplates the value of visualization in terms costs and gains. Our application of CMT to deconstruct a subset of rhetorical aspects of an infographic takes the view of visualization as an empirical science. Our approach is descriptive, and we work toward an understanding sufficient to formalize this theory in visualization practice. 

Visualization is pillared by empirical science that covers data, specification, task, or user-oriented tasks, all of which interrelate to varying degrees~\cite{van2005value,miksch2014matter,bruckner2018model}. A data-oriented perspective interrogates the data characteristics that are appropriate or possible to bring forward to a visualization. A specification-oriented perspective, in contrast, looks to the possible algorithms or parameters used to transform the visual encodings ~\cite{bertin1983semiology}. Task- and user-oriented perspectives~\cite{schulz2013design,brehmer2013} incorporate judgements of what encodings are \textit{effective} and \textit{expressive}, factoring in principles from perceptual and cognitive sciences~\cite{wertheimer1938laws,cleveland1984graphical,heer2010crowdsourcing} to produce legible and ``truthful'' depictions of data~\cite{tufte2001visual,callingbullshit2020}. Our work aligns most closely with task- and user-ori\-en\-ted ways of knowing about visualization, drawing from cognitive linguistics to explore how people relate to or understand unfamiliar content presented through the strategic use of visual conceptual metaphors.

\textbf{Conceptual metaphors in visualization.}
Leveraging metaphors to facilitate understanding is a frequent strategy in narrative visualization~\cite{segel2010narrative} and science infographic design~\cite{christiansen2022building} to communicate unfamiliar topics to broad audiences. Hullman and Diakopoulos~\cite{hullman2011visualization} include metaphors as a meaningful tool to evoke desired conceptual linkages or responses to the presented information. Their discussion of metaphors is relatively brief and high-level, while we take a deeper dive into conceptual metaphors applied specifically to scientific infographics. 

Although we frequently encounter visual metaphors, the scientific literature that provides design guidelines or evaluates conceptual metaphor use or value in understanding visualizations is limited. Tkachev et al.~\cite{tkachev2022metaphorical} apply conceptual metaphors to data, mapping unfamiliar data to a more familiar dataset to enable a facile exploration of relationships in the dataset. Our exploration lies at the level of visual encoding rather than a data-oriented perspective. 
Risch~\cite{risch2008role}, Cox~\cite{cox2006metaphoric}, and Parsons~\cite{parsons2018conceptual} review and discuss the role and use of conceptual metaphor as a communication tool in information visualization, with Parsons~\cite{parsons2018conceptual} providing a set of case studies to demonstrate the value of applying conceptual metaphor theory to the design and interpretation of such visualizations. Ziemkiewicz and Kosara~\cite{ziemkiewicz2008shaping,ziemkiewicz2009preconceptions} take these ideas a step further to explore how visual information can be shaped and differently perceived through a limited subset of conceptual metaphors: those of containment and levels. 
Our basic approach is grounded in these works, but we drill deeper into the different types of conceptual metaphors (the aforementioned \emph{imagistic}, \emph{orientational}, \emph{ontological}, and \emph{structural} metaphors) to show how these could be applied to visual scientific storytelling. 
Other work demonstrates the applicability of CMT to a specific domain. Sanchez et al.~\cite{sanchez2018visual}, explore the application of different classes of conceptual metaphors in medical photographs and graphics. Our work covers a different set of domains and instead focuses more narrowly on science infographics. Preim et al.~\cite{preim2023survey} comprehensively survey the use of metaphors for medical visualization interfaces, e.g., a ``surgical cockpit'' that borrows from aviation, while we analyze conceptual metaphors in visual assets.  

\section{Process}
Our process consisted of four main phases. 
First, we mapped the theoretical foundation of this work by \textbf{defining} conceptual metaphors and their subtypes in the context of visualization \cref{sec:define}. We then \textbf{selected} a corpus of science infographics for analysis based on a set of inclusion criteria (\cref{sec:select}), followed by \textbf{deconstruction and classification} of each infographic in the finalized corpus (\cref{sec:deconstruct}) and subsequent \textbf{synthesis} (\cref{sec:synthesize1}) and \textbf{summarization} (\cref{sec:synthesize2}) of our findings.

\subsection{Select corpus of science infographics}
\label{sec:select}

\begin{figure}[b]
        \centering 
         \includegraphics[width=\columnwidth]{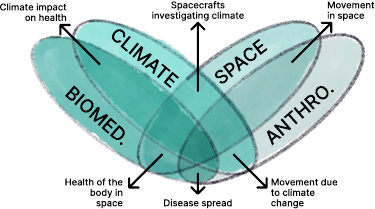}
         \caption{Illustration of the four scientific domains included in our corpus, represented by different saturations of teal. Their overlaps create a set of subdomains that are often found in public outreach. \textit{This representation is inspired by the graphic, ``What Makes a Good Visualization?'' \cite{mccandless2014knowledge}}.
         \label{fig:domains}}
    \end{figure}

\textbf{Domains.}
For our corpus, we selected four topic domains (biomedicine, climate, space, and anthropology) that frequently publish scientific infographics for the purpose of public engagement with science. We arrived at this set through initial exploratory web searches using the term ``science infographics'' and through extensive discussions amongst all coauthors of the topic domains we observed depicted in e.g., museums and popular science magazines. These four domains overlap and create combinations of sub-domains that also frequently appear in the context of public engagement, for instance, \emph{10: Graphical Abstract: Nano Silver-Induced Toxicity and Associated Mechanisms}, which primarily belongs to the biomedical domain and describes the effects on the body but is also relevant to the climate domain as it describes the effects of silver-induced toxicity from air pollution. We illustrate these domain overlaps in \cref{fig:domains}, with written examples of topics built on their intersection.

Within each domain, we restricted the scope to infographics artifacts that encompass various temporal and spatial ranges. We do not focus on a singular spatio-temporal range, as this could lead to a biased representation of metaphors in infographics. In the biomedical domain, for example, we include examples ranging from molecular processes (including cells, tissues, organism, etc.) to populations. In addition, the size or scope of the elements depicted in the infographics depends on the domain---the molecular scale in biomedicine being equivalent to local microclimates in the climate domain, space missions in the space domain, and local movements (e.g., neighborhoods) in anthropology.

\textbf{Inclusion criteria.} For inclusion in our study, an infographic:
    \begin{enumerate}[nosep,left=0pt .. \parindent]
        \item must be a scientific infographic, a visual representation of complex information aimed at novices to quickly explain the concepts by combining data with design \cite{smiciklas2012power};
        \item primarily uses graphical elements to convey the conceptual metaphor(s), and does not rely on text or photographs (although these may which be incorporated in and accompany the infographic);
        \item belongs to at least one of our chosen domains, selected for their frequent use of infographics in public outreach (e.g., data visualization websites, public health sites, online textbooks, etc.) describing concepts that can seem unfamiliar or abstract to people who are not domain experts (e.g., explanation of health, causes of global warming, stellar phenomena, migration of species over time, etc.) across a broad spatio-temporal range;
        \item is a static 2D representation such as a poster, leaflet, graphic abstract from a manuscript, or figure (\textit{not} an animation, video, interactive visualization, or aug\-men\-ted\discretionary{/}{}{/}vir\-tu\-al reality visualization) as a means to limit the scope of our analysis; 
        \item shows at least two entities that interact in a process, chosen through our observations that scientific processes tend to be rather abstract, complex phenomena that tend to use conceptual metaphors as an explanatory device; 
        \item does not present \textit{highly} visually complex phenomena (e.g., explaining several phenomena at the same time) that have the potential to introduce further risk of error or subjectivity into interpretation; and
        \item shows a spatio-temporal relationship since size and time are variables that can be measured and mapped when a process occurs. 
       \newline
    \end{enumerate}

\textbf{Search process \& keywords.}
We used a diverse range of outlets in academia and industry to find our corpus of infographics for analysis. These sources included data visualization pages, e.g., the \textit{Visual Capitalist}, \textit{Data Is Beautiful}, space agencies, e.g., \textit{National Aeronautics and Space Administration (NASA}), \textit{European Space Agency (ESA)}, graphical abstracts from academic papers relevant to our domains, e.g., \textit{Dovepress}, online history textbooks, e.g., \textit{Princeton Commons World History Encyclopedia}, and poster contest winners at biological and/or medical visualization conferences, e.g., Visual Computing for Biology and Medicine (VCBM) and Visualizing Biological Data (VIZBI). 

We adjusted our search words during the search process to achieve a thorough representation of the ranges of phenomena along the spatio-temporal dimensions. Primarily, we used keywords that related to the specific spatio-temporal scopes of the respective domains (see our \keywordslink{keywords document} for our initial keyword search). 
In the \emph{biomedical} domain, the smallest and fastest phenomena occur on the molecular scale (e.g., molecular reactions), and the largest affect whole populations (e.g., COVID-19 spread on a global scale). The shortest temporal occurrences are on a molecular level, and the most time-consuming evolve organs.
In the \emph{climate} domain, the smallest spatial changes occur on local levels (e.g., the effect of climate change on microclimates) compared to large-scale occurrences at the global level (global warming predictions). The least time-consuming changes take up to a year (e.g., fewer days with snow over a year), and the longest ones were naturally changing climates over millennia.  
In the \emph{space} domain, we searched for space missions by humans (which take up to a year and happen on a small scale relative to the size of the universe), followed by keywords based on planets, stars, galaxies, and the universe (i.e., larger and longer on the spatio-temporal scale). 
In \emph{anthropology}, the time periods range from phenomena of up to a year (e.g., migration during an armed conflict) to up to billions of years (e.g., human evolution from unicellular organisms). On the spatial scale, this domain covers local changes (e.g., boroughs in the city) to global movements all over the world. The analyzed infographics can be found in \databaselink{our database}.

\textbf{Corpus summary.}
Our initial search via keywords and organic search from seed infographics yielded an initial collection of 109 infographics, 55 of which passed our inclusion criteria. We have 10--15 infographics per domain, each of which we thoroughly deconstructed and classified to identify the types of metaphors in use. 



\subsection{Deconstruct \& classify each infographic}
\label{sec:deconstruct}
Drawing inspiration from linguistics, we adapted the Method for Identifying Metaphorically Used Words (MIP)~\cite{group2007mip}, an established approach for identifying textual metaphors in hierarchical lexical units of a text, to identify and classify visual conceptual metaphors. This was an iterative process, with initial classification done by one researcher and supported by weekly discussions with other members of the research team over a period of six months.  

We first focused on studying the infographic as a whole to identify the process that was being shown. We then deconstructed the infographic into its \emph{individual graphical units}, or graphical entities, that correspond to lexical units of a text \cite{group2007mip}. Several components can make up one \textit{graphical unit}, in cases where further division alters the context of the given graphical unit (e.g., per \cref{fig:teaser}, the symbol of a mitochondrion surrounded by coffee cups festooned with lightning bolts would lose its ontological meaning if the coffee cups were not considered within this unit). 
We then determined whether the graphical unit is a metaphor or not (e.g., an \emph{abstraction}), as detailed in \cref{conc_met}. If the graphical unit was a visual conceptual metaphor, we classified it as one of the four conceptual metaphors defined by CMT~\cite{lakoff1980metaphors}: \emph{imagistic}, \emph{orientational}, \emph{ontological}, or \emph{structural} (\cref{met_types}).


\subsection{Synthesize research artifacts}
\label{sec:synthesize1}

\begin{figure*}[tb]
        \centering 
         \includegraphics[width=\linewidth]{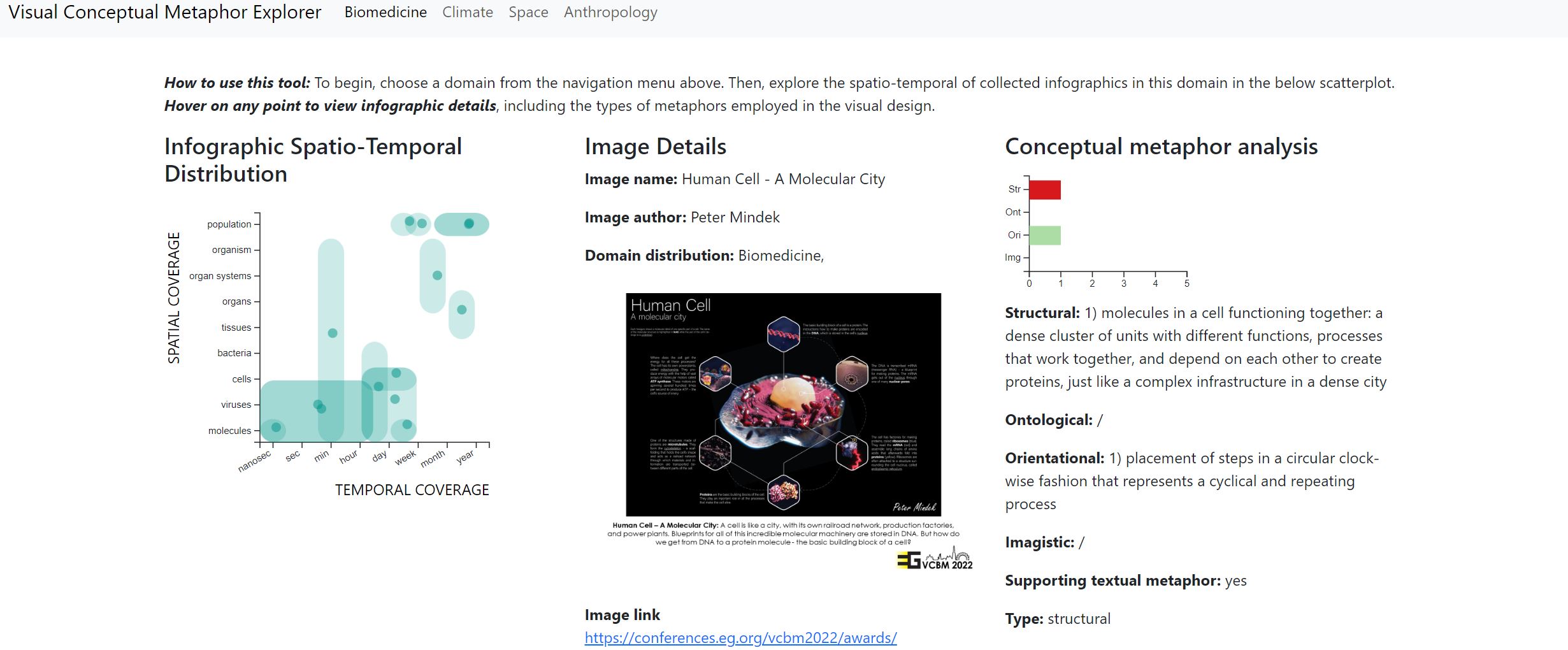}
         \caption{Screenshot of the Visual Conceptual Metaphor Explorer, developed to support the exploration of collected infographics and analyzed conceptual metaphors. For each of the four domains, a scatterplot depicts the spatial ($y$) and temporal ($x$) coverage of each infographic (left) and serves as the entry point to further exploration. Selection of a point displays the corresponding image and metadata (here, \textit{A Human Cell: A Molecular City} by Peter Mindek, reproduced with the author's permission) alongside our analysis of the visual conceptual metaphors present (right).}
         \label{fig:exploratoryTool}
    \end{figure*}
    
The classification of each metaphor in our corpus yielded two primary research artifacts: 
(1) a \databaselink{database of classified infographics} and (2) a \toollink{visual exploratory tool}, available \osflink{on OSF}\footnote{\osflinkrepo}. 
These supported our synthesis of the occurrences, patterns, and implications of using different types of conceptual metaphors in the design of scientific infographics.

\textbf{Conceptual Metaphor Database.}
To demonstrate our classification process and inspired by the VIS30K Collection \cite{chen2021vis30k}, we created a \databaselink{database} with the final 55 analyzed and coded infographics. This database allowed us to observe some repeating patterns in the infographics, e.g., linguistic metaphors in the text to support graphical conceptual metaphors, and contains the following information:
\begin{itemize}[nosep,left=0pt .. \parindent]
    \item \emph{General information about the infographic}: infographic title, author (s), the domain(s) it belongs to, an embedded icon, a URL link, and an image URL address for easier viewing. 
    \item \emph{Brief overview of the properties}: target audience, the cultural background, and the primary communication goal for making the infographic determined by checking the source from which we retrieved the infographic (e.g., Center for Disease Control and Prevention), the visual style (e.g., illustrative rendering), or a subjectively determined level of complexity used to explain the science.   
    \item \emph{Classification of the conceptual metaphor(s)}: a Boolean statement for the presence of each and a description of how they are presented.
    \item \emph{Other observations}: use of affective visualization, the use of insets, whether the infographic shows occurrences across scales, whether the visualization displays comparisons between elements it visualizes, and whether it shows a cyclical process. We also recorded whether the text is present in the infographic and, if so, whether it supports the visual, conceptual metaphors. 
    \item \emph{Check for criteria fulfillment}: to note whether the infographics depict a process, we noted the number and names of entity types and whether they show a process by interacting.
    \item{\emph{Axis placement}}: notion of categorizing the infographics on the spatio-temporal axis ($x = temporal coverage, y = spatial coverage$).
    \item {\emph{Usage rights}}: license type (if available) and copyright document. 
\end{itemize}

\textbf{Visual Conceptual Metaphor Explorer.}
An \toollink{exploratory tool} built using \texttt{D3.js} accompanies this database, which enables visual inspection of our corpus of classified infographics. 
This tool allows users to explore the spatio-temporal range spanned by each infographic for each domain, and to observe patterns in visual conceptual metaphor distribution. 
We divide the interface into three sections (\cref{fig:exploratoryTool}): spatio-temporal distribution of all infographics in a given domain with each infographic plotted as a rectangular shape with a circle centroid (left), general information for the selected infographic (middle), and the analysis of the visual conceptual metaphors within the chosen infographic (right). From the domain spatio-temporal distribution chart, the user selects a circle centroid to display the chosen infographic with its title, author, and source URL link in the middle panel. Simultaneously, we show a bar chart visualization of the conceptual visual metaphors within the selected infographic in the analysis section (right). We chose this idiom as it clearly and quickly displays the count of metaphor types present in each infographic. Below the bar chart, we include further details on each analyzed visual metaphor as well as if and how text supported the visual metaphors of the infographic.

\subsection{Summarize \& reflect on metaphor corpus}
\label{sec:synthesize2}

\begin{table}[tb]
        \centering 
         \caption{Total number of visual conceptual metaphor types, of related textual metaphors, and of the use of effective visualization per domain.}%
         \label{tab:domains}%
         \includegraphics[width=\columnwidth]{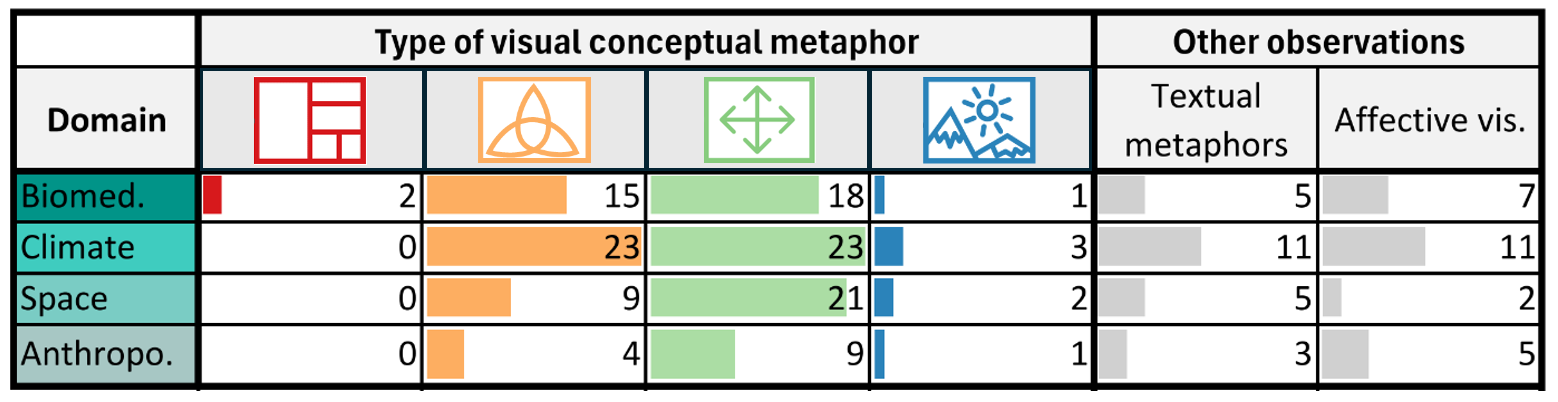}%
\end{table}

In \autoref{tab:domains} we summarize the frequencies of each type of conceptual metaphor across the four domains that we analyzed in our corpus. We observed a total of 131 conceptual metaphors across all 55 infographics (\emph{structural}$=$2, \emph{ontological}$=$51, \emph{orientational}$=$71, \emph{imagistic}$=$7).  
We further synthesize the results of our deconstruction and classification process here, continuing with a discussion of possible explanations and further reflections on study findings in \cref{discuss}.


\textbf{Limited metaphors in anthropology.}
Compared to the other domains, anthropology used few conceptual metaphors. This domain instead relied on visual abstractions such as arrows to represent time and movement, visual marks like larger/ thicker arrows to encode amounts of people, and glyphs to label important locations of cities and battles (e.g., \emph{48: Migration Period in Europe, 5th Century}; \cref{fig:infographics_2}c). Similarly, visual abstractions in the form of colors were generally used to label and distinguish different groups of people or geographical areas without any semantic meaning.

\textbf{Reliance on ontological \& orientational metaphors.}
We found that \emph{structural} and \emph{imagistic} metaphors were not used as extensively as the other two types. \emph{Orientational} (n$=$71) and \emph{ontological} (n$=$51) conceptual metaphors were used the most across domains, with \emph{orientational} metaphors dominating all four domains. The space and anthropology domains relied primarily on \emph{orientational}  metaphors, biomedicine mostly on \emph{orientational} but also heavily on \emph{ontological} metaphors, while climate used both types equally. 

\emph{Orientational} metaphors tended to support the communication of cyclical processes with different design strategies, e.g., 
arranging stages of a process into a circle (e.g., \emph{3: A Human Cell: A Molecular City}, also depicted in \cref{fig:exploratoryTool}) or an infinity symbol (e.g., \emph{1: Metaphor and novelty in science education visuals}, shown in \cref{fig:teaser}). 
\emph{Orientational} metaphors also often depicted time, where usually \textit{left} represented the past and \textit{right} the future. The space domain, however, had several exceptions, e.g., representing time as a spiral (e.g., \emph{29: Earth Fleet}) or time passing from right to left, as if observing stellar phenomena that occurred myriad years ago and which are only now observable (\emph{34: Development of Massive Elliptical Galaxies}).

\textbf{Affective strategies with ontological \& orientational metaphors.}
Affective visualization, mainly using \emph{ontological} conceptual metaphors, was primarily used in the climate domain (n=11). This strategy frequently represented the effects of global warming, how warming can be alleviated (e.g., \emph{20: Keeping It Cool}), and its consequences in the present day (e.g., \emph{18: State of the UK Climate 2020}, shown in \cref{fig:infographics}b) and future (e.g., \emph{16: What is Climate Change?}). The most common techniques relied on \emph{ontological} metaphors, where colors red, orange, or yellow represented high temperatures alongside a sense of urgency, often coupled with text calling for action (e.g., \emph{22: Western Wildfires and Climate Change}, seen in \cref{fig:infographics_2}b). 
Another tactic used an \emph{orientational} metaphor that compared the present and future through juxtaposition. Similarly, infographics from several domains, such as biomedicine (n$=$7) integrated affective visualization through \emph{ontological} and \emph{orientational} conceptual metaphors to visualize disease prevention (\emph{5: How Protein Subunits of COVID-19 Vaccines Work}). 
Anthropology used affective visualization almost as frequently as the biomedical domain (n$=$5), e.g., when it overlapped with the biomedical domain and described disease spread and urgency (\emph{46: Containment and Health Index (CHI) by Country on the First Day of Each Month}), armed conflict (\emph{45: Europe's New Migration Crisis}), or negative impact on economy for a given country (\emph{52: Millionaire Migrations 2023}). 
The space domain used affective visualization the least (n$=$2), as this domain mainly focused on stating facts and making complex ideas more understandable. An example of affective visualization is \emph{31: Journey to Mars}, which aimed to popularize science by evoking excitement and a sense of adventure.

\textbf{Text to support visual metaphors.}
Sometimes textual metaphors were used to support graphical conceptual metaphors within the infographics, which we observed most often in the climate domain (n$=$11), followed by biomedicine (n$=$7), anthropology (n$=$5), and was used the least in space (n$=$3). The anthropology domain offers an example in the \emph{52: Columbian Exchange}, where the title itself helps the reader understand the origin and unfair trade between the continents. The text supports the visual metaphors in the form of deliberate color choices (\emph{ontological}) for the diseases passed on from overseas.  

\section{Discussion}
\label{discuss}
We explored the feasibility of classifying visual conceptual metaphors in science infographics in an effort to formalize the process of metaphor use in infographic and visualization more broadly. Next, we expand on patterns and themes observed over the course of our investigation. 

\begin{figure}[tb]
        \centering 
         \includegraphics[width=\columnwidth]{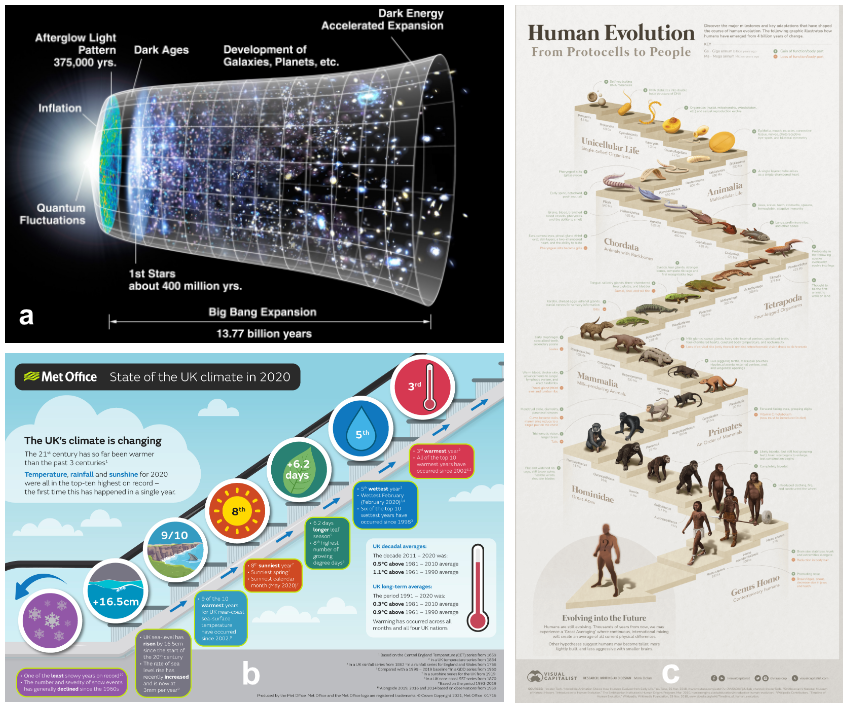}\vspace{-1ex}
         \caption{Examples of the discussed infographics: (a) shows how the combination of \emph{ontological} and \emph{orientational} metaphors creates an optimal visual explanation for an abstract idea (expansion of the universe), (b) displays an infographic where the text `warmest day' compliments a semantically used red color \emph{ontological} that together refer to hot temperature, (c) shows an infographic where the conceptual metaphors can have several explanations based on a viewer's understanding of the content.}%
         
         \label{fig:infographics}
    \end{figure}

\textbf{Orientational \& ontological metaphors may be optimal balance for visual complexity and clarity.}
While Lakoff and Johnson \cite{lakoff1980metaphors} identified structural metaphors to be the most frequent type used in linguistics, we did not make the same observation for the use of visual conceptual metaphors in scientific infographics. 
We reason that the use of \emph{ontological} and \emph{orientational} metaphors in place of the most complex \emph{structural} metaphors may be strategically chosen on some level to break down scientific concepts into simpler, more easily-read conceptual metaphors. We suspect that this choice results in part from the scientific infographic visual medium itself. Many infographics are created to engage and be memorable to the viewer \cite{borkin2013makes}, which can require additional viewer cognitive effort and time than text alone \cite{ottley2019text}. To avoid overloading the user, the designer may, consciously or not, reach for visual conceptual metaphors that are themselves less complex while clearly communicating the intended message. A deeper exploration of this is an interesting direction for future work. 
%
An example is \emph{17: CO$_{2}$ Emissions vs. Vulnerability to Climate Change by Nation}, in which the countries with the biggest CO$_{2}$ production are positioned against and above the countries most negatively affected by CO$_{2}$, which take the lower part of the infographics. This visualization creates a juxtaposition between the causes and effects of climate change through visual composition and elicits tension and conflict through color: the red color for the emitters draws attention and creates a sense of urgency. This graphic does not go further (nor does it need to) to explain in detail how carbon dioxide affects the countries, the health of its individuals, or global warming, and instead focuses its rhetoric on the imbalance that needs to be addressed. 
Another example, \emph{30: The Big Bang and Universe Expansion} (\cref{fig:infographics}a), shows the use of an \emph{ontological} metaphor as it represents an abstract, ever-expanding space in all directions (the universe) as a cone that would be otherwise difficult to imagine. By setting imaginary boundaries through containment, the reader can imagine the main concept about the universe's expansion through time (represented by an \emph{orientational} metaphor, with left meaning the past and right meaning the future) without needing to understand the underlying complex astrophysical mechanisms. 
%


\textbf{Conceptual metaphors use reflects the needs and challenges of a given domain.}
Different domains leverage visual conceptual metaphors in different ways. An interesting occurrence is the portrayal of time, heavily influenced by a Western culture where it is a norm to read from left to right \cite{lakoff1980metaphors}, as we observed in all four domains. The space domain, however, used a variety of \emph{orientational} conceptual metaphors to describe the passing of time, e.g., spirals or an orientation from right to left. This use of `nonstandard' orientational metaphors results from explaining something that is not yet fully understood or culturally ingrained. For most people who are not researching astronomy or physics, the entry point to understanding these concepts may be achieved best through metaphor. Extra creativity on the designer's part may be needed to craft metaphors sufficient to describe the phenomena and that spark public engagement---this opens many possibilities for the development and investigation of new, effective strategies for visual conceptual metaphors. 

\begin{figure}[tb]
        \centering 
         \includegraphics[width=\columnwidth]{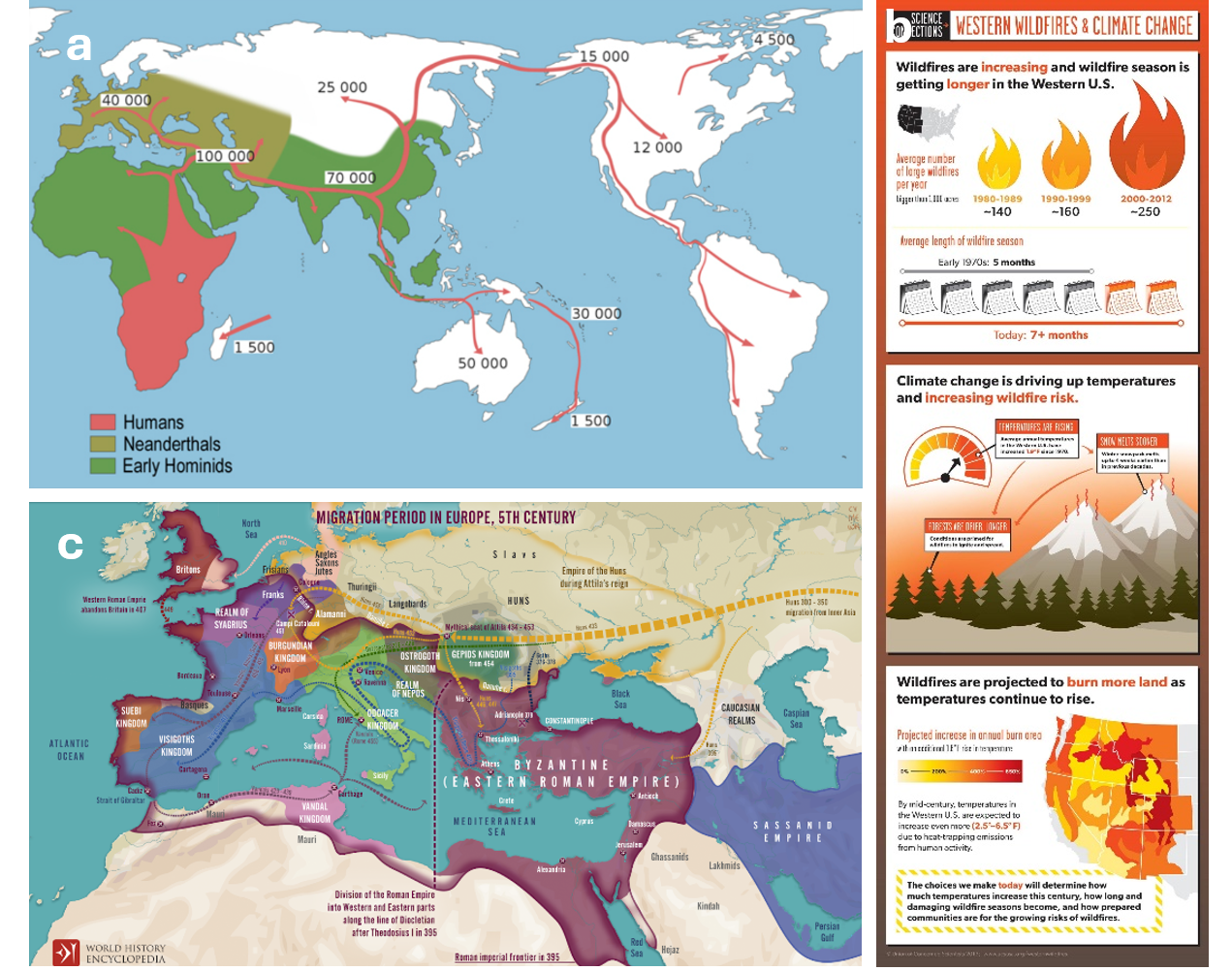}\vspace{-1ex}
         \caption{Examples of infographic showing (a) a map that utilizes minimalism and only one type of metaphor (orientational) to illustrate the idea of human migration, (b) a climate infographic utilizing ontological metaphor (red color to represent urgency) to create an affect, and (c) displays a map that relies on symbolism such as arrows and glyphs (abstractions) to represent fifth-century European battles and capital cities.}%
         
         \label{fig:infographics_2}
    \end{figure}

The anthropology domain used conceptual visual metaphors the least. We attribute this pattern (specifically in maps) to placing importance on displaying data with more accuracy and minimalism, similar to cartography. In terms of describing movement through time, e.g., \emph{44: Human Migration Map} (\cref{fig:infographics_2}a), more complex mapping is not necessary to get the main message across. In cases like these, visual abstractions suffice since the main actors are distinguished by color (that does not carry any meaning) and arrows to show movement over continents. Both concepts are thus represented as visual abstractions, not conceptual metaphors. 
While not included in our corpus for failing to meet our inclusion criteria, it is noteworthy that earlier cartographic visualizations (e.g., \emph{103: Map of Iceland}) employed visual conceptual metaphors of sea monsters (\emph{ontological}) that represented the dangers of unexplored waters. It is an interesting observation specific to this domain that shows that, after people understood the reason behind dangerous nautical areas, the concept of something scary represented by monsters (\emph{ontological}) was replaced by visual abstraction of the actual cause of the problem, e.g., shallow water. This example shows that visual conceptual metaphors helped viewers to derive information (source: danger) \cite{lakoff1980metaphors} and, through the use of effective visualization \cite{samsel2021affective}, act a certain way \cite{xiao2021we}. This aspect offers insight into how cognition extends its frontiers \cite{fauconnier1997mappings} of reasoning through image schemas because, even though the reason for dangerous seas was not understood, people knew to avoid it. In contrast today, people in the same profession know to avoid dangerous areas without having to reason through fear.
 
\textbf{Metaphors may promote more affective visualization.}
Another interesting phenomenon was the role of \emph{ontological} and \emph{orientational} conceptual metaphors in affective visualization \cite{lan2023affective} prevalent in the climate domain. \emph{Ontological} metaphors, such as colors, can be subtly manipulated to influence emotions and learning \cite{plass2014emotional}. To achieve this, semantic colors representing rising temperature and urgency simultaneously \cite{lan2023affective, samsel2021affective} can grab the reader's attention and engage them emotionally. 
These infographics often included a call to action through text \cite{hullman2011visualization}, as seen in \emph{22: Western Wildfires and Climate Change} (\cref{fig:infographics_2}b), which relies on a red color (\emph{ontological} metaphor representing urgency and notes how choices we make today will impact our future. We discuss the role of text and how it can support an affective message carried by the visual metaphor in further detail below. 
%
While composition does not automatically produce an \textit{orientational} metaphor, we observed instances where composition created an \textit{orientational} metaphor designed to elicit viewer urgency for change and action, 
e.g., the left versus right composition of a normal versus dystopian future inciting an \textit{orientational} metaphor through our notion of the linear passage of time: \textit{left}, the past, is normal while \textit{right}, the future, is bleak. 

The affective use of visual conceptual metaphors was presented mainly in the climate domain, where a large number of people are needed to take action against something by playing a part in the prevention or living through the consequences. 
Another example of affective visualization can be seen in the biomedical domain, where groups or individuals are responsible for their own health as well as disease spread to other people. In both of these cases, affective strategies paired with conceptual metaphors target the viewers' emotions to subliminally persuade them to take action. A deeper investigation into the implications on engagement and response to a visual's call to action is a promising avenue for future work.

\textbf{Text and visual conceptual metaphors often cooperate.}
Following previous research \cite{hullman2011visualization} on the value of accompanying text (e.g., titles, captions) to visual representations, text helped to scaffold the meaning of the graphic metaphors within the infographic. 
In \emph{18: State of the UK Climate 2020} (\cref{fig:infographics}b), the temperature increase was represented in text by stating the ``warmest day'' placed next to a red field where color semantically and intuitively represented ``hot'' temperature (\emph{ontological}---and was used to create a sense of urgency. At the same time, the \emph{orientational} metaphor was used to place the visual element at the top of the infographics representing increase. This example shows multiple encodings for a given visual conceptual metaphor, where text further supports the conceptual metaphor displayed in the graphical unit. 
%
We furthermore see instances where a conceptual metaphor in text enhances or changes the interpretation of a visual conceptual metaphor when considered as a unit. For instance, \cref{fig:teaser} includes text describing the cell as a `package.' This textual \emph{ontological} metaphor lends additional meaning to the \emph{imagistic} metaphor for cellular membrane, represented as a stitched patch that forms the outer boundary of the cell and contains the text positioned within it.

Text can be an equalizer for people coming from different levels of expertise and can aid in pinpointing the context. Another benefit of using text in combination with graphics is that it may help viewers distinguish between different interpretations of metaphor types depicted in a given visualization to improve the clarity of the overall message. We leave a further investigation on the relationship between textual and visual conceptual metaphor use and practice as future work. 

\textbf{Boundaries may be blurred for classification \& interpretation.}
The interpretation of certain visual conceptual metaphors is not always clear, as their classification boundaries can be blurry. Sometimes, they clash and almost seem counter-intuitive. 

For example, \emph{54: Visualized: The 4 Billion Year Path of Human Evolution} (\cref{fig:infographics}c) displays human evolution through biological entities going down steps, where we note two possible interpretations that depend on the viewer's subjective experiences and subsequent reasoning about the phenomenon in question. Observed through the lens of an \emph{ontological} metaphor, one reads the downward stair progression as an instantiation of the phylogenetic tree from distant to recent taxonomic ancestry. This interpretation requires some degree of knowledge and belief in scientific theory. An alternative \emph{orientational} metaphoric lens may interpret the downward orientation of evolution as a decline or other negative insinuation. Equally, the visual design of the stairs may not have been intended by the designer as a metaphor at all and may depict an Escher-like\cite{escher1960ascending} illusion illustration for purely visual effect. This case is an example where a metaphor's interpretation is scaffolded in part by the viewer's previous knowledge and familiarity \cite{lakoff1980metaphors} with evolutionary taxonomy, highlighting the impact of prior knowledge. In that case, visualizations aimed at a certain group of people ought to be designed in a way that is tailored to that group \cite{raz2019epigenetic, pokojna2023seeing}. This observation ties into previous research demonstrating that the viewer's thinking style impacts how they interpret the metaphor \cite{ziemkiewicz2010understanding}. This case is representative of many different interpretations that each facilitate understanding of complex and abstract concepts through personal experience and familiar associations.

Another example of a blurred categorization occurs in \emph{7: Increasing Physical Activity Among Adults with Disabilities}, where the patient's recovery is outlined as a guide for the physician through a winding, discretized path \emph{ontological}. One could argue that this metaphor is \emph{structural} as it can be represented as a \textit{road to recovery} with many steps (a process where the patient is subjected to appointments, consultations, evaluations, and discipline that takes time and potential setbacks), compared to a journey that is also a long process that entails many different steps, time, and setbacks (\emph{structural}). Conversely, the interpretation we chose after extensive discussion was an \emph{ontological} interpretation, which focuses on the graphic from a physician's perspective, where the graphical unit of the path as procedural steps that do not encompass the larger picture of various facets of illness and recuperation. 

Likewise, \emph{19: State of UK Climate 2022} illustrates the temperature rise through a Newton's Cradle metaphor to depict cause and effect. At first glance it may use a \emph{structural} metaphor to represent a complex set of causes and effects of climate change. It once again, however, serves primarily as an \emph{ontological} metaphor, where the power of the metaphor conveys the idea of a sequence of causes and effects in global warming without additional metaphors to build a deeper network of meaning. 


Finally, it may be tempting to overthink a graphical representation and apply more conceptual meaning than it possesses. The following examples refer to the shortlisted, yet excluded, subcollection of infographics in our \databaselink{database}. \emph{11: The Many Phases of Silver} shows a war axe cutting a bacterium in half to represent destruction of bacteria with silver. This may seem like an \emph{imagistic} metaphor. However, the war axe itself is a symbol for killing or destruction, and is furthermore colored silver. Since the source and target fall in the same conceptual domain, this representation is a visual \textit{abstraction}.  
%
As another example, consider \emph{48: Migration Period in Europe, 5\textsuperscript{th} Century} (displayed in \cref{fig:infographics_2}c), which marks important sites such as battles with glyphs of two crossed swords that are not conceptually different from a battle, making them a visual abstraction. The infographic \emph{50: 1346--1353 spread of the Black Death in Europe map} makes use of different types of arrows (dashed and filled) that represent different means of movement (maritime and land trade routes), but these graphics do not cross to a different conceptual domain to represent this notion. Ultimately, not all infographics need a metaphor to convey a concept---if the entity or process at hand is sufficiently familiar, an abstraction may be able to convey the necessary information without the extra cognitive effort required to interpret a metaphor. 
We furthermore encourage the visualization research community to engage in deep reads of infographics to help disambiguate abstractions from metaphors.

\textbf{Imprecise use of \textit{metaphor} in visualization.}
Related to our discussion above, an overarching point that we observed in our process of researching and assembling science infographics for our analysis is the lack of precision in the use of the term \textit{metaphor} in visualization. We observed that \textit{metaphor} can refer to \textit{abstraction}, \textit{symbolism}, or any sort of visual representation of an otherwise abstract entity. For instance, MetaGlyph~\cite{Ying2023MetaGlyph} is a project that proposes to automatically generate glyphs that serve as suitable \textit{visual metaphors} to represent relevant underlying dataset semantics. Here, the authors' use of the term \textit{metaphor} aligns more with abstraction rather than a conceptual metaphor. We believe a more differentiated use of the term within the visualization community would be beneficial for disambiguating between these related but distinct concepts. 

\section{Limitations \& future work}
Our classification and meta-discussion of types of visual conceptual metaphors in science infographics should serve as the starting point to understand the intentional and informed use of conceptual metaphors in visualization. The exploration of visual conceptual metaphors unveils new avenues within the narrative visualization field that can be explored. Yet, our analysis and methodology are not without limitations.
    
In sourcing infographics (and due to searching via English keywords) we mostly found Western examples. As discussed in previous work \cite{lakoff1980metaphors, hullman2011visualization}, cultural, philosophical, and religious beliefs influence the understanding and appreciation of conceptual metaphors.
We saw this aspect, e.g., in the depiction of time in our corpus, which is mostly represented through the linear, horizontal orientational metaphor of the left as the past and the representing the future. Such representations may differ in cultures that read from right to left. Considering other \emph{orientational} metaphors from different cultural perspectives, religious groups may negatively interpret striving `upwards' towards the divine as an act of hubris. Collectivist cultures \cite{triandis2001individualism} may find resonance with a central (good)---peripheral (negative) orientation \cite{lakoff1980metaphors}. 
Similarly, the orientation and composition of visual elements can be interpreted differently based on a biological basis, as shown by studies that show left-handed people perceive the left side as positive, while right-handed people perceive this in the opposite way \cite{casasanto2009embodiment}. Cultural ingraining can also influence the semantic encoding of certain elements, e.g., the color blue in Eastern culture symbolizes inflammation while red which is used to indicate the same phenomenon in Western cultures \cite{kitao1986study}. An analogous exploration of our work that analyzes visuals beyond Western culture may help illuminate the standard application of visual conceptual metaphors.

Our study is also limited in only looking at four domains, which we chose based on the infographics most likely to be seen due to their importance and popularity as scientific topics discussed in public: educational institutions (e.g., schools, textbooks), public outreach materials (e.g., museums, doctor's offices, news, maps), or visualization research dissemination (conferences, research abstracts). As we saw, different domains take advantage of different visual and conceptual metaphors. Exploring domains beyond biomedicine, climate, space, and anthropology would thus be beneficial in future work that expands this study of CMT applicability and use in other domains. This would help towards the establishment of an eventual taxonomy of visual conceptual metaphors across multiple disciplines, e.g., in politics groups \cite{oushakine2016translating}.

Our investigation targeted visual representations generally for a lay audience, and we did not explore visualization types beyond infographics. The next step should thus expand this horizon into different areas of visualization and visual communication through different sources curated to varying target groups, e.g., newspapers or children's books.

We see our feasibility study and meta-discussion on the application of CMT to classify types of conceptual metaphors in science infographics as laying the groundwork for developing a visual taxonomy and grammar applicable to the broader visualization field, building further on work such as by  Ziemkiewicz and Kosara~\cite{ziemkiewicz2008shaping,ziemkiewicz2009preconceptions} and on theories and perspectives for visualization design \cite{bederson2003craft,van2005value,chen2017pathways} and interaction \cite{beaudouin2004designing}. As part of this endeavor, further evaluation of the extensibility of and edge cases in our approach is necessary. 
Achieving such a formalization may aid in establishing a procedure for creating and evaluating visualizations using visual conceptual metaphors in the vein of Kindlmann's algebraic approach to visualization \cite{kindlmann2014algebraic}. 
Future lines of investigation could build upon our study of CMT types applied to infographics to explore semi- or even fully-automated strategies for producing the right type of conceptual metaphor to convey the intended idea, similar to previous work with glyphs \cite{Ying2023MetaGlyph} or via prompts for the ever-evolving generative AI imagery systems.
Further user studies would shed light on mapping the conceptual thinking behind abstract scientific occurrences. An interesting area to explore would be evaluating outcomes between AI-generated infographics, trained scientific illustrators, and computer scientists. 
An automated approach to the graphic display of metaphorical information may ensure greater objectivity in data visualization by omitting human manipulation and employing algorithms to standardize layout, formatting, and data representation based on attributes of the dataset \cite{mackinlay1986automating}. Visual metaphors going beyond semiology, e.g., in illustrative visualization, where a reusable template based on datasets' properties \cite{sorger2017metamorphers} was already explored and could be an inspiration for this direction. Finally, we note that while such conceivable approaches would likely be time-effective to produce, the risk of such automated or generative processes is the diminishing of a human creative design perspective, which may not always be desirable or beneficial. 

\section{Conclusion}
Metaphors serve us as a means of communicating the unfamiliar. Metaphors are ubiquitous, visible in myriad data journalism articles, academic literature, and more broadly in popular culture. Their use is not always explicit, as in the cartoon television series ``Once Upon a Time\dots Life'' \cite{barille1987}, which personifies various complex biological processes (an ontological metaphor) to make science approachable to younger audiences. We argue that awareness of these types of conceptual metaphors and an understanding of their patterns of use have the potential to facilitate design processes for science infographics that are ultimately more effective in their communication aims. 
Our work represents the first step to evaluating types of visual metaphors in scientific infographics,  creating a basis that could be developed for taxonomy and framework to create infographics effectively. 
Through this study, we have furthermore deepened the understanding of visual conceptual metaphors' application by deconstructing and classifying metaphors of existing science infographics that could be applied to the visualization field as a whole and contribute to improving our visual vocabulary.





\acknowledgments{
The authors wish to thank Matúš Talčík for his help in developing the Visual Conceptual Metaphor Explorer. This work is supported in part by the University of Bergen and the Trond Mohn Foundation in Bergen (\#813558, Visualizing Data Science for Large Scale Hypothesis Management in Imaging Biomarker Discovery (VIDI) Project). 
}

\label{sec:appendices_inst}


\section*{Supplemental materials}
\label{sec:supplemental_materials}
All supplementary materials for this work are available \osflink{on OSF} (\osflinkrepo) and include:
\begin{enumerate}[nosep,left=0pt .. \parindent]
    \item the \databaselink{\emph{database}} of analyzed infographics and infographics that were shortlisted but excluded from the final analysis for failing to meet all inclusion criteria,
    \item our \toollink{\emph{exploratory tool}} visualizing infographics distribution and visual conceptual metaphor analysis, which contains the \githubrepolink{GitHub repository URL} and \livetoollink{live tool link} in the README document,
    \item \keywordslink{\emph{keywords}} we used in initial searches for infographics,
    \item the \figureslink{\emph{figures} and \emph{table}} used in this paper. 
\end{enumerate}

\section*{Figure credits}
\label{sec:figure_credits}

\Cref{fig:teaser} is a partial recreation of the poster contest winner at VIZBI Conference (March 28--31, 2023) \cite{VIZBI2023}, used under the CC BY-NC-ND license obtained from the author by written agreement.
\Cref{fig:exploratoryTool} depicts the People's Choice Award image from VCBM Workshop 2022 (September 22--23, 2022) by Peter Mindek, from \cite{VCBM2022}, under CC-BY license obtained from the author by written agreement.
\Cref{fig:infographics} displays (a)\emph{\bigbanglink{The Big Bang and Universe Expansion}} by NASA/WMAP Science Team -- Original version: NASA; modified by Cherkash obtained from Wikipedia, which is in the public domain, (b) \emph{\stateoftheuklink{State of the UK Climate 2020}} by the Royal Meteorological Society for educational purposes and with written permission, and (c) \emph{\humanevolutionlink{Visualized: The 4 billion-year path of human evolution}} by Mark Belan from Visual Capitalist, used for individual research. \cref{fig:infographics_2} shows (a) \emph{\humanmigrationmaplink{44: Human Migration Map}} from the Migration Heritage NSW for the purpose of research under Copyright Act, (b) \emph{\migrationperiodlink{Migration Period in Europe, 5th Century}} by Simeon Netchev
from \worldhistoryencyclopedialink{World History Encyclopedia} under the Creative Commons Attribution-NonCommercial-ShareAlike license, and (c) \emph{\wildfireslink{Western Wildfires and Climate Change}} made by the Union of Concerned Scientists for educational purposes.

For all remaining {\emph{figures} and \emph{tables}, we, as authors, state that these are and remain under our own personal copyright, with permission to be used here. We also make them available under the \href{https://creativecommons.org/licenses/by/4.0/}{Creative Commons At\-tri\-bu\-tion 4.0 International (\ccLogo\,\ccAttribution\ \mbox{CC BY 4.0})} license and share them at \osflinkrepo.

\bibliographystyle{abbrv-doi-hyperref}

\bibliography{abbreviations,template}



\end{document}